\pdfoutput=1
\documentclass[superscriptaddress, 12pt, aps, prl,  preprint]{revtex4-1}
\usepackage{graphicx}
\usepackage{epstopdf}
\usepackage{color}
\usepackage{amsmath, amssymb}

\begin{document}

\title{Dielectric properties of condensed systems composed of fragments}
\author{Ding Pan}
\email{dingpan@ust.hk}
\affiliation{Department of Physics and Department of Chemistry, Hong Kong University of Science and Technology, Hong Kong, China}
\affiliation{HKUST Fok Ying Tung Research Institute, Guangzhou, China}
\author{Marco Govoni}
\affiliation{Materials Science Division and Institute for Molecular Engineering, Argonne National Laboratory, Argonne, Illinois 60439, USA}
\affiliation{Institute for Molecular Engineering, The University of Chicago, Chicago, Illinois 60637, USA}
\author{Giulia Galli}
\affiliation{Institute for Molecular Engineering, The University of Chicago, Chicago, Illinois 60637, USA}
\affiliation{Department of Chemistry, The University of Chicago, Chicago, Illinois 60637, USA}
\affiliation{Materials Science Division and Institute for Molecular Engineering, Argonne National Laboratory, Argonne, Illinois 60439, USA}

\date{\today}

\begin{abstract}
The dielectric properties of molecules or nanostructures are usually modified in a complex manner, when assembled into a condensed phase. We propose a first-principles method to compute polarizabilities of sub-entities of solids and liquids, which accounts for multipolar interactions at all orders, and is applicable to any semiconductor or insulator.  The method only requires the evaluation of induced fields in the condensed phase, with no need of multiple calculations for each constituent. As an example, we present results for the molecular polarizabilities of water in a wide pressure and temperature range. We found that at ambient conditions, the dipole-induced-dipole approximation is sufficiently accurate and the Clausius-Mossotti relation may be used, e.g. to obtain molecular polarizabilities from experimental refractive indexes. However with increasing pressure this approximation becomes unreliable and in the case of ice X the Clausius-Mossotti relation is not valid.
\end{abstract}

\maketitle

\section{introduction}

The polarizability of molecules and nano-structures is an  important  property  determining the assembly and behavior of solids and liquids composed of  well-defined building blocks. In addition, molecular polarizabilities play a key role in vibrational spectroscopy, e.g. in the calculation of Raman \cite{putrino2002anharmonic} and sum-frequency generation \cite{wan2015first} spectra, in the determination of 
van der Waals interactions in solids and liquids \cite{mahan1982van, klimevs2012perspective}, and in the development of polarizable force fields.

While several methods are available to compute and predict polarizabilities of isolated molecules and nanostructures, their definition and calculation in condensed phases have been challenging  and various levels of approximations have been adopted in the literature (e.g., \cite{feynman, kittel2004, heaton2006condensed, salanne2008polarizabilities, wan2013raman}). For example, the Clausius-Mossotti (CM) equation relates the average atomic or molecular polarizability $\alpha$  of a material building block to its  electronic dielectric constant  $\epsilon_\infty$\cite{feynman,kittel2004}: 
\begin{equation}\label{CM}
  \frac{4\pi N\alpha}{3}=\frac{\epsilon_\infty-1}{\epsilon_\infty+2},
\end{equation}
where $N$ is the number density of atoms or molecules. In Eq. \ref{CM} if we substitute $\epsilon_\infty$ with the refractive index of the material $n$  using $\epsilon_\infty=n^2$,   the Lorentz-Lorentz equation is recovered.  The validity of the CM relation in condensed phases, such as molecular liquids or assembly of nanostructured solids,   depends on the system and general rules to establish its regime of applicability are not available.  We note that often times the variation of polarizabilities from the gas to the condensed phase are neglected. For example, many molecular dynamics simulations of aqueous solutions using force fields assume a fixed molecular polarizability of water \cite{schropp2008polarizability}, though first-principles electronic structure studies of  water at ambient conditions have shown that molecular polarizabilities have rather broad distributions and are not isotropic \cite{heaton2006condensed, salanne2008polarizabilities, wan2013raman}. Most electronic structure methods \cite{heaton2006condensed, salanne2008polarizabilities, wan2013raman} consider only the dipole-dipole interaction when calculating the variation of polarizabilities upon assembly of molecular fluids or solids, and in many cases such approximations have remained untested. Recently, substantial progress  has been reported  in computing polarizabilities of building blocks using maximally localized Wannier functions (MLWFs)\cite{marzari2012maximally}. However, this method requires separate calculations of the dielectric properties of each constituent self-consistently, and it is not based on global induced fields within the condensed system \cite{PhysRevB.92.241107,PhysRevB.96.075114}.

In this letter, we propose a first-principles method to compute the polarizabilities of building blocks in condensed phases. The method, bases solely on electronic structure calculations for the condensed phase, is  applicable to any semiconductor or insulator. We present results for the molecular polarizablities of water in a wide pressure-temperature (P-T) range, and we validate the CM relation for water at ambient conditions and the dipole-induced-dipole approximation (DID). We found that the DID becomes increasingly less accurate under pressure and breaks down when covalent bonds are present and oxygen ions are formed within the solid.

We start by summarizing our formulation. A building block (BB) composing a condensed system (e.g. a molecule in a molecular crystal)  is defined by its ionic coordinates  and by  electronic wave functions spatially localized at the BB site, for example  maximally localized Wannier functions $w_i(\vec{r})$\cite{marzari2012maximally} constructed from the  Bloch orbitals of the condensed phase.  The linearly induced electron polarization density of the BB, in response to a macroscopic field $\vec{E}$ is:
\begin{equation} \label{mlwf}
 \Delta \rho_{BB}(\vec{r}) =  2 \sum_i^{N_{orb}} w_i^*(\vec{r})\Delta w_i(\vec{r}) + c.c.,
\end{equation} 
where $N_{orb}$ is the number of localized electronic orbitals (e.g. four doubly-degenerate orbitals for a water molecule with 8 valence electrons), and $\Delta w_i$ is the variation of the $i$-th Wannier function.   The local  field ($\vec{E}_{loc}$) acting on the BB is given by two contributions:
\begin{equation}\label{E_loc}
\vec{E}_{loc} = \vec{E} + \vec{E}_{env},
\end{equation}
where  $\vec{E}_{env}$ denotes the field produced by the environment surrounding the BB, that is by all the electrons that do not belong to the BB. In most  previous studies, $\vec{E}_{env}$ was approximated by dipole-induced-dipole (DID) electrostatic interactions\cite{heaton2006condensed, salanne2008polarizabilities, wan2013raman}. 

The polarizability tensor $\alpha_{BB}$ of the BB is defined by the equation
\begin{equation}\label{mol_pol}
        \vec{\mu}_{BB} = \alpha_{BB}\vec{E}_{loc}, 
\end{equation}
where 
$\vec{\mu}_{BB}$ is the dipole moment computed as:
\begin{equation} \label{mu}
\vec{\mu}_{BB} = -q_e\int \vec{r} \Delta \rho_{BB}(\vec{r})d\vec{r}.
\end{equation} 
In Eq. (\ref{mu}),  $q_e$ is the elementary charge and $\Delta \rho_{BB}$ is from Eq. (\ref{mlwf}).
To compute $\alpha_{BB}$, we need to calculate $\vec{E}_{loc}$. Since $\vec{E}$ is fixed, we only need to determine $\vec{E}_{env}$, which is simply: 
\begin{equation}\label{projector}
\vec{E}_{env} = 
\frac{1}{N_{orb}} \sum_i^{N_{orb}} \int \vec{e'}(\vec{r}) \left| w_i(\vec{r}) \right|^2 d\vec{r}  \,,
\end{equation}
where  $\vec{e'}(\vec{r})$ is the \emph{microscopic} electric field induced by all the electrons outside the BB. 

The microscopic electric field $\vec{e'}$ can be evaluated  within the  
random phase approximation (RPA) or by including the variation of the exchange and correlation potential (we denote the latter with DFT). Within RPA, $\vec{e'}_{RPA}$ is obtained using Gauss' law \footnote{Historically, it is from the \textit{wing} part of the inverse dielectric matrix\cite{baldereschi1979microscopic,PhysRevB.23.6615}}:
\begin{equation} \label{rpa}
\nabla \cdot \vec{e'}_{RPA}(\vec{r}) = -4\pi q_e \Delta \rho'(\vec{r}),
\end{equation} 
where $\Delta \rho' = \Delta \rho - \Delta \rho_{BB}$ and $\Delta \rho$ is the electron polarization density of the whole system.
At the DFT level, the exchange-correlation potential $V_{xc}$ also contributes to the microscopic local field:
\begin{equation} \label{dft}
\vec{e'}_{DFT}(\vec{r}) =  \vec{e'}_{RPA}(\vec{r}) + \frac{1}{q_e} \nabla \left(\frac{d V_{xc}}{d\rho} \Delta \rho' (\vec{r})\right).
\end{equation}

Once $E_{loc}$ and $\mu$ are computed from Eqs. (\ref{E_loc}) and (\ref{mu}), respectively, $\alpha_{BB}$ is known. 
Therefore the procedure outlined here to obtain polarizabilities of BB within a condensed system is rather simple. Once the density and single particle wavefunctions are computed, e.g. by solving the Kohn-Sham equations, the electron polarization density is obtained by performing a single self-consistent calculation for the whole system.  Eq. (\ref{rpa}) or (\ref{dft}) are then solved non-self-consistently, including multiple interactions at all orders. Solvers to obtain linear variations of the electron density exist in most DFT codes, using either density-functional pertubation theory (DFPT) \cite{baroni2001phonons} or finite fields \cite{PhysRevLett.89.117602}.

We now turn to applying the method outlined above to the study of 
the molecular polarizabilities of water in a broad P-T range from ambient to supercritical conditions.
The electron polarization density was obtained by DFPT, as implemented in the plane-wave pseudopotential code Qbox (http://qboxcode.org/) \cite{gygi2008architecture,wan2013raman} 
\footnote{We used Hamann-Schluter-Chiang-Vanderbilt norm-conserving pseudopotentials \cite{PhysRevLett.43.1494,PhysRevB.32.8412} with a plane-wave kinetic energy cutoff of 85 Ry. The MD trajectories of water at ambient conditions were taken from the water PBE400 dataset \cite{gygiPBE400}: http://www.quantum-simulation.org, where there are 64 water molecules in the simulation box. The supercritical water trajectories were from our previous simulations \cite{pan2013dielectric,pan2014refractive},  where the simulation box has 128 water molecules. The MD trajectories of the Na$^+$-water solution is from Ref. \cite{gaiduk2017local}. At least 60 snapshots from each MD trajectory were employed in our electronic structure calculations. 
For ice VIII and ice X, we used a 96 and 128-molecule supercells, respectively; the results were validated using  a Monkhorst-Pack k-point mesh of 8$\times$8$\times$8 with the primitive cells \cite{monkhorst1976special}.}.
The Perdew-Burke-Ernzerhof (PBE) exchange-correlation (xc) functional \cite{PhysRevLett.77.3865} was used. Although 
PBE overestimates the molecular polarizability of an isolated water molecule by $\sim$10\%, for water under pressure PBE gives both the static and the electronic dielectric constants in better agreement with experimental values than at ambient conditions, as shown in previous studies \cite{pan2013dielectric,pan2014refractive}. Here we used one xc functional to analyze trends of polarizabilities as a function of P and T, however the method is general and can be used with any functional. In particular we note that using finite field methods to compute polarizabilities (http://qboxcode.org/)\cite{wan2013raman}, calculations with hybrid functionals are readily carried out.

\begin{figure}
\centering
\vspace{5mm}
\includegraphics[width=0.6 \textwidth]{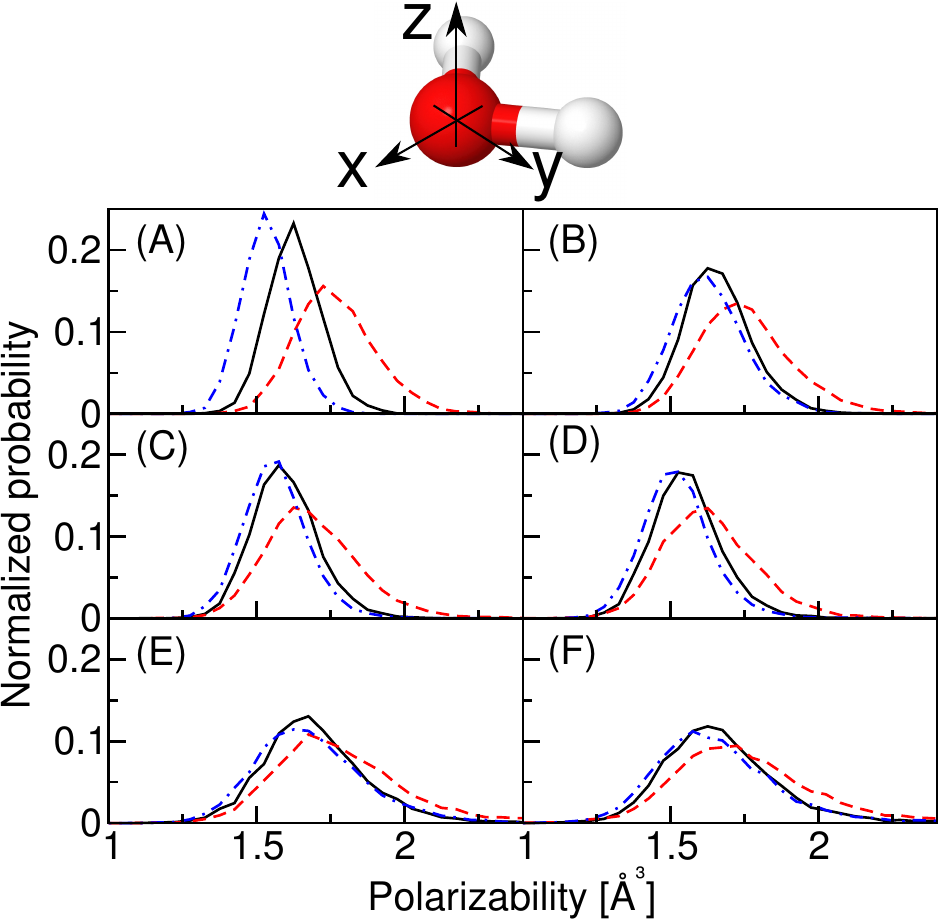}
\caption{Distribution of three diagonal elements of polarizability tensors of water molecules at various pressure-temperature conditions. x: along the bisector of $\angle$HOH; y: in the molecular plane and perpendicular to x; z: out of the molecular plane. Solid black, dash-dotted blue, and dashed red lines represent the xx, yy, and zz elements of polarizability tensors respectively. (A) the ambient condition; (B) 1000 K, 1.1 GPa; (C) 1000 K, 5.8 GPa; (D) 1000 K, 11.4 GPa; (E) 2000 K, 5.2 GPa; (F) 2000 K, 8.9 GPa. }
\label{distriDFT}
\end{figure}

Fig. \ref{distriDFT} shows that at ambient conditions, the molecular polarizabilities of water given by the DFT method (Eq. (\ref{dft})) are anisotropic.  The out of plane polarizability is the largest, and the ones in-plane and perpendicular to the water dipole direction are smaller, consistent with the reports of other authors using just DID interactions to compute $E_{env} $\cite{heaton2006condensed, salanne2008polarizabilities, wan2013raman}. 
We found that at high pressures and high temperatures, the anisotropy substantially decreases as shown in Fig. \ref{distriDFT}. 
Note that an isolated water molecule also exhibits a  polarizability which is less anisotropic than in the liquid at ambient conditions \cite{wan2013raman,PhysRevB.96.075114},  
suggesting that the anisotropy is critically related to the formation of hydrogen bonds. Indeed, also in supercritical water, the polarizability components are less dissimilar than at ambient conditions (see  Fig. \ref{distriDFT}). 
 
In Table \ref{table}, four different methods to compute polarizabilites are compared 
 from ambient to 11 GPa and 0 to 2000 K. 
All methods show that with increasing pressure along an isotherm, the average molecular polarizability of water ($\bar{\alpha}_{mol}=\frac{1}{3}\mathrm{Tr} \{\alpha_{mol}\}$) decreases,
while with increasing temperature along an isobar, it increases.
Our previous study showed that the average dipole moment of water molecules increases with pressure, 
but decreases with temperature, 
so the present results indicate that varying the molecular dipole moments of water becomes more difficult when the values of the moments increase.

The polarizabilities obtained by DFT (Eq. (\ref{dft})) are slightly larger than
those from RPA (Eq. (\ref{rpa})) by $\sim$0.02 \AA$^3$. 
When applying the two methods, we used the same electron polarization density $\Delta \rho$, which is obtained when the exchange-correlation functional is included. The local electric field mainly comes from the electrostatic interactions, 
so the DFT and RPA values are very similar.

In order to test the validity of the CM relation, we substituted the electronic dielectric constant $\epsilon_\infty$, obtained by DFPT into the CM relation to calculate the average molecular polarizability. 
It is interesting to see that the CM relation yields nearly identical results as the DID approximation.
The standard deviations obtained for the CM relation are smaller than those from the DID approximation by one order of magnitude, 
as they only arise from the thermal fluctuation of $\epsilon_\infty$,  not from  molecular distributions as shown in Fig. \ref{distriDFT}.
The CM relation holds when the Lorentz relation holds in an isotropic material, that is to say that the field at the center of a fictitious spherical cavity created by molecules inside the cavity vanishes \cite{kittel2004}.
A well-known example for the Lorentz relation is the lattice with cubic symmetry, where only dipole-dipole interactions are considered \cite{kittel2004}.
The agreement between the results obtained using the CM relation and the DID approximation suggests that the Lorentz relation is accurate when we consider only the dipole-dipole interaction for the water systems studied in Table \ref{table}.

We now turn to comparing molecular polarizabilities of water in various phases.
If we substitute the refractive index of 1.333, the experimental value for water at 293 K and ambient pressure, into the CM relation, 
we get a molecular polarizability of 1.47 \AA$^3$, which is the same as the experimental value for water vapor.
At the PBE level of theory, the polarizability of an isolated water molecule is 1.60 \AA$^3$, the same as the values obtained at ambient conditions using the CM relation and DID approximation (see Table \ref{table}),  consistent with previous studies \cite{wan2013raman}.
Hence, within the DID approximation, the average molecular polarizability of water does not change from gas to liquid phase.
However, both the RPA and DFT methods give slightly larger values ($\sim$3\% than that obtained with CM and DID methods). 
We note that recently, Ge and Lu reported the molecular polarizabilities of water and ice at ambient conditions calculated using the local dielectric response of orbitals \cite{PhysRevB.96.075114}, where the electron polarization density of each molecule $\Delta \rho_{mol}$ is evaluated individually. In general the sum of $\Delta \rho_{mol}$ does not equate the total $\Delta \rho$. In calculations of   Ref.\cite{PhysRevB.96.075114}, the $\bar{\alpha}_{mol}$ of water increases by $\sim$10\% (instead of $\sim$3\%) from gas to the liquid at ambient conditions. 

Using the DFT method, we also calculated the molecular polarizabilities of water in the first solvation shell of the Na$^+$ ion at ambient conditions\cite{gaiduk2017local}: $\bar{\alpha}_{mol}$ is 1.62 \AA$^3$, which is again slightly larger than that obtained by the DID approximation  by $\sim$3\%. For water molecules with dangling bonds in the basal surface layer of ice Ih \cite{pan2008surface, watkins2011large}, the difference in $\bar{\alpha}_{mol}$ given by the DFT and DID approaches is even smaller: 1.62 \AA$^3$ vs 1.61 \AA$^3$, only $\sim$1\%. Our results suggest that at ambient conditions, the DID approximation works remarkably well.

\begin{figure}
\centering
\vspace{5mm}
\includegraphics[width=0.6 \textwidth]{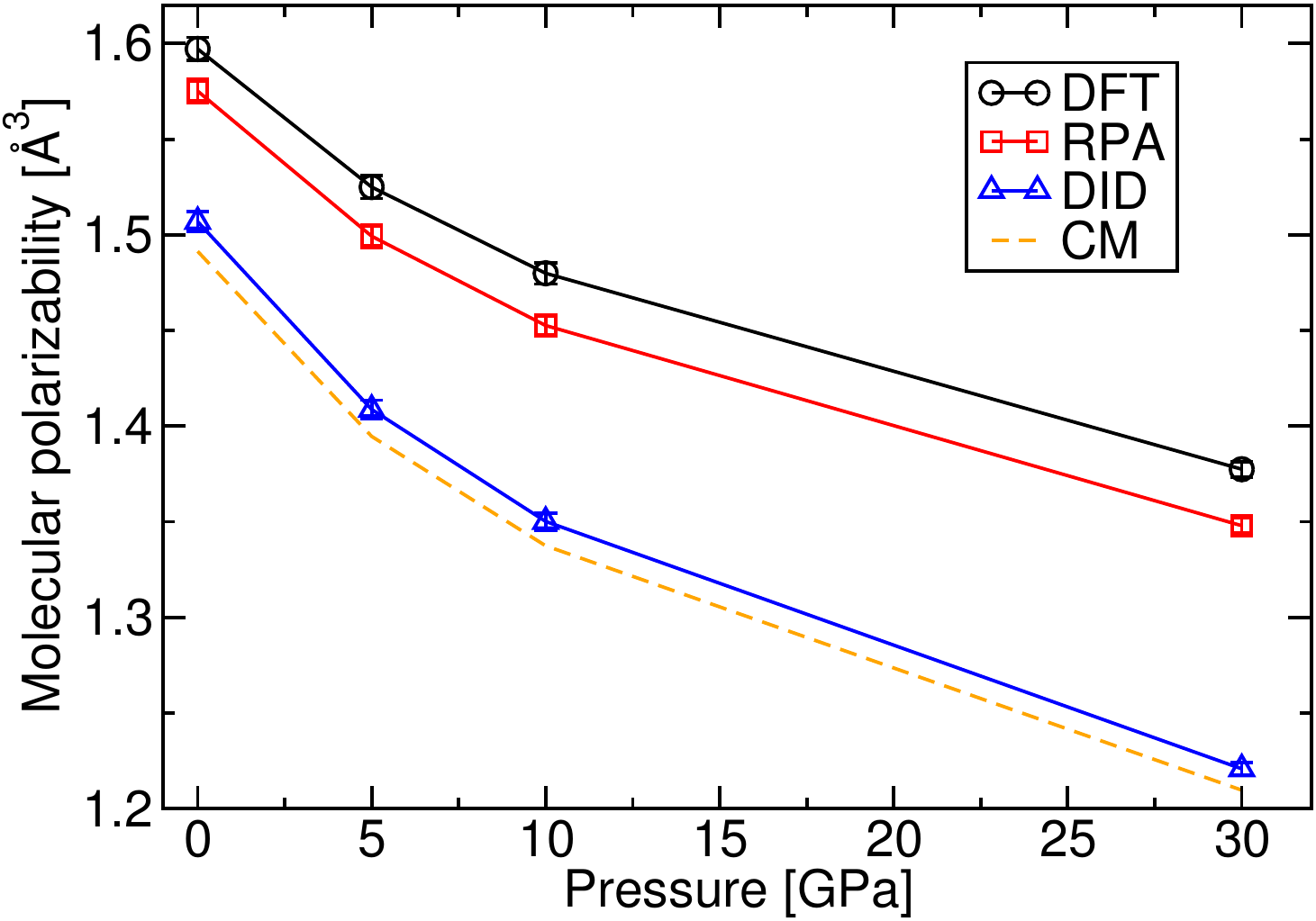}
\caption{The molecular polarizabilities of ice VIII under pressure. The results obtained by four methods are shown:
density functional theory (DFT), random phase approximation (RPA), dipole-induced-dipole (DID), and Clausius-Mossotti (CM) relation.}
\label{iceVIII-pol}
\end{figure}

Table \ref{table} shows that  the polarizabilities  obtained by our method and the CM relation or the DID approximation differ  with increasing pressure at a fixed temperature. For ice VIII, a high pressure ice phase consisting of two interpenetrating cubic ice sublattices \cite{petrenko1999physics}, when increasing pressure from 0 to 30 GPa, the difference between  DFT values and the DID approximation increases from 6\% to 13\%, as shown in Fig. \ref{iceVIII-pol}. It indicates that  interactions higher than dipole-dipole play a bigger role for  denser water. 

\begin{figure}
\centering
\vspace{5mm}
\includegraphics[width=0.6 \textwidth]{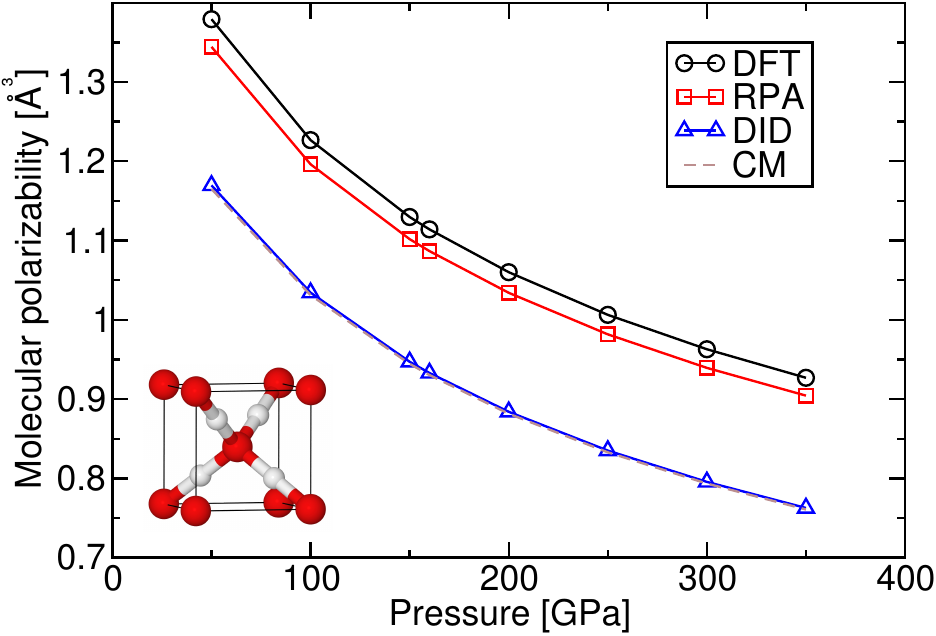}
\caption{The molecular polarizabilities of ice X under high pressure. The results obtained by four methods are shown:
density functional theory (DFT), random phase approximation (RPA), dipole-induced-dipole (DID), and Clausius-Mossotti (CM) relation. The DID and CM results are indistinguishable.}
\label{iceX-pol}
\end{figure}

We close by considering the case of extremely dense water: ice X, the highest pressure phase ever determined experimentally \cite{hemley1987static}. In ice X, the oxygen atoms are in a body-centered cubic lattice, and the hydrogen atoms sit right between two nearest O atoms (see Fig. \ref{iceX-pol}). Because the H atom is equidistant to two O atoms, it is no longer possible to define H$_2$O molecules; however since the four maximally localized Wannier orbitals are still closely localized around O atoms, a new BB can be defined, and the molecular polarizability discussed below refers to the polarizability of the O$^{2-}$ anion.

Fig. \ref{iceX-pol} shows that the CM relation gives the same results as the DID approximation 
whereas the molecular polarizabilities given by the DFT and RPA methods are about 20\% larger. The reason is that in ice X  covalent bonds are present, and indeed the BB identified by our calculation is no longer a water molecule, but rather an anion, for which  higher-order interactions play an important role.

\begin{figure}
\centering
\vspace{5mm}
\includegraphics[width=0.6 \textwidth]{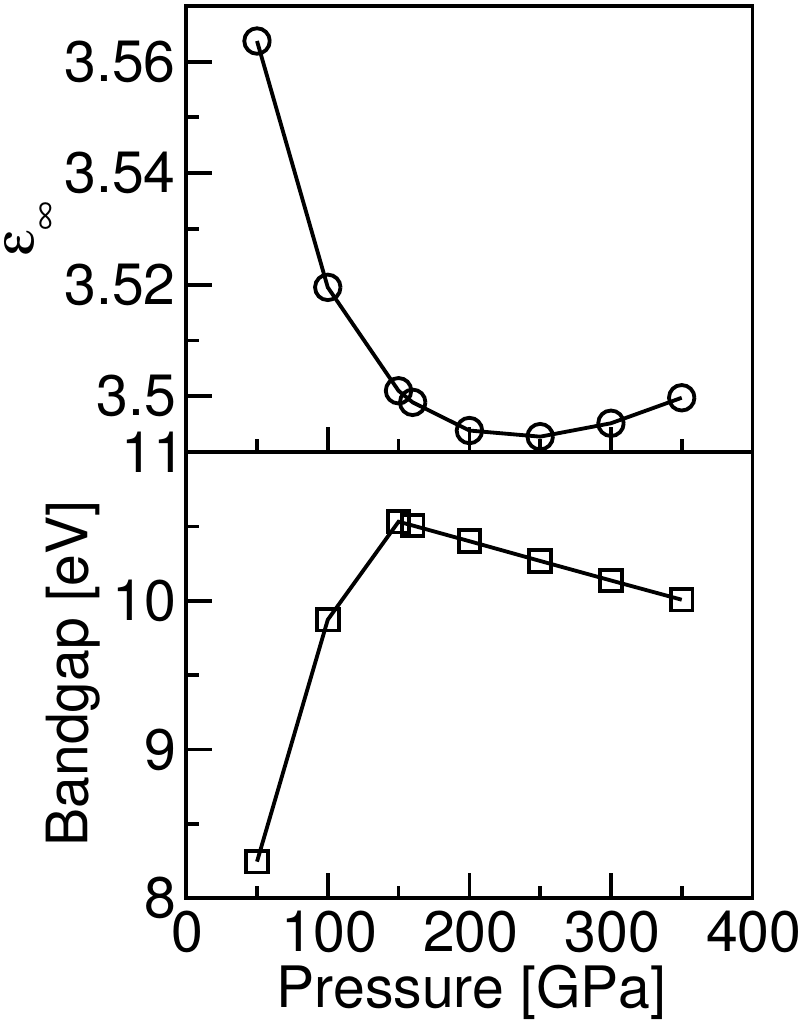}
\caption{Upper panel: the electronic dielectric constant of ice X as a function of pressure; Lower panel: the band gap of ice X as a function of pressure.}
\label{epsi-gap}
\end{figure}

For ice X, another interesting finding of our calculation is that the electronic dielectric constant $\epsilon_\infty$ has a minimum at around 250 GPa, and
accordingly the band gap increases up to 150 GPa and then decreases slowly, as shown in Fig. \ref{epsi-gap}. The inverse correlation between the electronic dielectric constant and the band gap of ice X is consistent with the Penn model \cite{PhysRev.128.2093, angilella2017correlations},  and differs from what we found in ice VII/VIII and  hot water up to 30 GPa in our previous study \cite{pan2014refractive}. 
Generally, $\epsilon_\infty$ increases when both molecular polarizability and material density become larger. With increasing pressure, the molecular polarizability of ice X decreases as shown in Fig. \ref{iceX-pol}, whereas the material density increases due to volume shrinking, so the molecular polarizability and the material density of ice X are two competing factors determining  $\epsilon_\infty$; this is also the reason why the variation of $\epsilon_\infty$ is weak (see Fig. \ref{epsi-gap}).
From 50 to 250 GPa, the molecular polarizability dominates the change of $\epsilon_\infty$, but above 250 GPa, the rate of its decrease becomes slower and thus the material density becomes a more important factor. As a result, $\epsilon_\infty$ decreases slowly as shown in Fig. \ref{epsi-gap}.

\section{Conclusion}
In order to predict the properties of solids and liquids composed of well defined building blocks, it is important to determine the variation of the dielectric properties of the isolated molecular or nano-scale constituents upon assembly. Hence the ability to compute dipole moments and polarizabilities of building blocks in condensed phases is critical. In this paper  we proposed a first-principles method to compute polarizabilities of sub-entities in condensed phases, which includes  multipole interactions at all orders and is applicable to any semiconductor or insulator. The methods only requires a single self-consistent calculation for the entire condensed system, as opposed to multiple calculations for each building block, and it is readily applicable within and beyond the RPA. As an example, we presented results for the molecular polarizabilities of liquid water in a wide pressure and temperature range. We found that at ambient conditions, the dipole-induced-dipole approximation is sufficiently accurate and the Clausius-Mossotti relation may be used, e.g. to obtain molecular polarizabilities from experimental refractive indexes. However with increasing pressure this approximation becomes unreliable and in the case of ice X, the Clausius-Mossotti relation is not valid. Interestingly, we found that the DID is increasingly less accurate under pressure. 
For example in ice VIII the contribution of multipole beyond the dipole is $\sim$13\% at 30 GPa and in ice X, the difference between  all multiple  and the DID contribution  is about  $\sim$20\% at 350 GPa, 
indicating that when hydrogen bonds are replaced by covalent bonds,  higher-order interactions cannot be ignored. 
In the case of ice X the CM relation is not valid, though the Lorentz relation still holds under the DID approximation. 
We also found that the band gap of ice X has a maximum, while the electronic dielectric constant of ice X has a minimum, as a function of pressure. Finally we note that 
the knowledge of the polarizabilities of sub-entities under pressure may help to design polarizable force fields suitable for extreme P-T conditions. 
 The method presented here can be used to study the local dielectric response of a wide range of semiconductors and insulators, and brings new insights into chemical bond interactions.

\section{acknowledgements}
We thank Deyu Lu, He Ma, and Ikutaro Hamada for their helpful discussions.
 D.P. acknowledges support from Hong Kong Research Grants Council (project number ECS-26305017), the National Natural Science Foundation of China (project number 11774072), the Alfred P. Sloan Foundation through the Deep Carbon Observatory, and the Croucher Foundation through the Croucher Innovation Grant.
M.G. and G.G. were supported by MICCoM, as part of the Computational Materials Sciences Program funded by the U.S. Department of Energy, Office of Science, Basic Energy Sciences, Materials Sciences and Engineering Division. This research used resources of the Research Computing Center at the University of Chicago, the Argonne Leadership Computing Facility, which is a DOE Office of Science User Facility supported under contract DE-AC02-06CH11357.

\bibliography{ref}
 
\newpage

\begin{table}
\centering
\caption{The average value of the three diagonal elements of polarizability tensors of water molecules obtained by four methods:
Clausius-Mossotti (CM) relation, dipole-induced-dipole (DID), random phase approximation (RPA), and density functional theory (DFT). 
The standard deviations of data in the molecular dynamics simulations are shown in parentheses. (Unit: \AA$^3$)}
\label{table}
\begin{tabular}{|l|c|c|c|c|}
\hline
 & CM & DID & RPA & DFT \\ \hline
          ambient & 1.605 (0.007) & 1.595 (0.082)   & 1.625 (0.082)   & 1.643 (0.083)  \\ \hline
1.1 GPa, 1000 K & 1.638 (0.009) & 1.626 (0.107)  & 1.668 (0.109)  & 1.681 (0.110) \\ \hline
5.8 GPa, 1000 K & 1.526 (0.009) & 1.521 (0.107)  & 1.599 (0.109)  & 1.619 (0.111)  \\ \hline
11.4 GPa, 1000 K & 1.453 (0.009) &  1.449 (0.109)  & 1.549 (0.114) & 1.572 (0.116) \\ \hline
5.2 GPa, 2000 K & 1.639 (0.017) & 1.628 (0.187)   &  1.712 (0.201) &  1.730 (0.204)\\ \hline
8.9 GPa, 2000 K  & 1.586 (0.022) &  1.578 (0.251) & 1.688 (0.273) & 1.711 (0.276)\\ \hline
1st solvation shell of Na$^+$ & - & 1.569 (0.140) & 1.607 (0.133) & 1.623 (0.130) \\ \hline
ice Ih \{0001\} surface & - & 1.610 (0.114) & 1.607 (0.120) & 1.622 (0.114)\\ \hline 
\end{tabular}
\end{table}

\end{document}